\documentstyle[aps,prl,multicol,epsfig]{revtex}

\begin{document}
\title {Inversion of K$_3$C$_{60}$ Reflectance Data}
\def\bea {\begin{eqnarray}}
\def\eea {\end{eqnarray}}
\def\be {\begin{equation}}
\def\ee {\end{equation}}
\def\eqnum#1{\eqno (#1)}
\def\sup{superconductivity }
\def\high{high $T_c$ superconductivity }
\author{F. Marsiglio$^{1,3}$, T. Startseva$^2$, and J.P. Carbotte$^{2,3}$}
\address {$^1$Department of Physics, University of Alberta, Edmonton, AB 
T6G 2J1\\
$^2$Dept. of Physics \& Astronomy, McMaster University, Hamilton, Ontario 
L8S 4M1 \\
$^3 $ Canadian Institute for Advanced Research, McMaster University,
Hamilton, ON L8S 4M1}

\maketitle
\begin{abstract}
We outline a procedure for obtaining the electron-phonon spectral density
by inversion of optical conductivity data, a process very similar in spirit
to the McMillan-Rowell inversion of tunelling data. We assume both
electron-impurity (elastic) and electron-phonon (inelastic)
scattering processes. This procedure has the advantage that it can be
utilized in the normal state. Furthermore, a very good qualitative result
can be obtained {\em explicitly}, without iteration. We illustrate this
technique on recently acquired far-infrared data in K$_3$C$_{60}$. We
show that the electron-phonon interaction is most likely responsible for
superconductivity in these materials.
\end{abstract}
\begin{multicols}{2}
The determination of the underlying interactions and how they
govern the various phases in materials (metallic, insulating, 
superconducting, etc.) is a major goal in current research of new
materials. In the last ten years, both high-T$_c$ oxides and alkali-doped
``bucky-balls'' (A$_3$C$_{60}$, etc.) have provided fascinating cases
for study, since several phases are attainable by both doping and
temperature variation. The general problem of understanding how these
various phases come about is a formidable task, however, and remains unsolved.

In this paper we investigate the origin of superconductivity in
K$_3$C$_{60}$, a superconducting fullerene with $T_c \approx 19$ K.
While several experimental facts point towards a conventional
electron-phonon mechanism (eg. isotope effect measurements
\cite{chen,ramirez} and others --- see Refs. \cite{gelfand}, \cite{pickett},
\cite{pennington} and \cite{gunnarsson}
for reviews), the key experiment,
which is single electron tunneling at biases above the gap edge,
has not yet been performed \cite{ostrick} to the extent that reproducible
structure is observed. In fact, definitive results from single electron
tunneling may be very difficult to achieve for two reasons. First, the
electron-phonon coupling strength may be very weak, in which case the structure
will be difficult to observe, and second, neutron scattering experiments
indicate that phonon modes are present at very high frequencies
\cite{pintschovius} ($\omega > 200$ meV). If these modes are coupled to
the itinerant electrons, then they may be impossible to see through
tunneling experiments, whose accuracy becomes less reliable at higher
biases.

For these reasons we attempt here to provide as definitive a demonstration
of electron-phonon superconductivity as tunneling normally provides, using,
however, a higher frequency and potentially more sensitive probe, infrared
spectroscopy. This idea is certainly not new, as many years ago Allen
\cite{allen71} provided a rudimentary basis for inferring the electron-phonon
spectrum for lead, based on the absorption measurements of Joyce
and Richards \cite{joyce}. This work was followed up by
Farnworth and Timusk \cite{farnworth}; they utilized weak coupling expressions
due to Allen and were able to extract a spectrum which agreed very well
with that obtained from tunneling.

The present work differs somewhat from these approaches in that we
utilize only normal state reflectance measurements, over the (in principle)
entire frequency range. Through a Kramers-Kronig transformation we
obtain the optical conductivity, from which we can deduce the
electron-phonon spectrum. We first present
some of the theoretical aspects, carry out some checks on Pb, and finally
apply our procedure to K$_3$C$_{60}$. Our conclusion is that the optical
measurements of Degiorgi et al. \cite{degiorgi} are consistent with
electron-phonon coupling of sufficient strength to drive the
superconducting transition in K$_3$C$_{60}$.

The starting point is the Kubo formula for the optical conductivity
\cite{mahan} of electrons in a normal metal, at zero temperature:
\be
\sigma(\nu) = {\omega^2_P \over 4\pi} {i \over \nu}
\int_0^\nu d\omega {1 \over \nu + {i \over \tau} - \Sigma(\omega)
- \Sigma(\nu - \omega) }
\label{sigma}
\ee
\noindent where
\be
\Sigma(\omega) = \int_0^\infty d\Omega \alpha^2F(\Omega)
\ln | {\Omega - \omega \over \Omega + \omega} | - i\pi
\int_0^{|\omega|} d \Omega \ \alpha^2F(\Omega)
\label{selfenergy}
\ee
\noindent is the electron self-energy due to the electron-phonon
interaction (the term proportional to the electron-impurity scattering
rate is explicitly included in Eq. (\ref{sigma})). Eq. (\ref{sigma})
has been derived using a number of well-documented
approximations: impurity scattering
is included in the simplest Born approximation, and electron-phonon
vertex corrections have been ignored. Actually we have written the
self-energy as a functional of $\alpha^2F(\Omega)$, although Allen
\cite{allen71} has noted (following Scher \cite{scher}) that this form
can be used provided $\alpha^2F(\Omega)$ is replaced by a more complicated
spectral function called $\alpha^2_{ir}F(\Omega)$, which in some limits
is very well approximated by  $\alpha^2_{tr}F(\Omega)$. This latter function
contains the essential ``$1 - cos(\theta)$'' factor for transport properties.
So, in fact the above equations should contain the transport spectral function,
but since this and the usual spectral function are very often
qualitatively similar, we have dropped the distinction.

It should be clear from Eqs. (\ref{sigma}) and (\ref{selfenergy})
that the normal state conductivity contains an image of the
underlying electron-phonon spectral function. This is most apparent
if we invoke a weak coupling approximation and expand the denominator
in Eq. (\ref{sigma}) to first order in the self-energy, so one of the
integrals can be performed by parts \cite{remark_allen}. Further 
straightforward manipulation leads to the following {\it explicit} 
expression for the electron-phonon spectral function:
\be
\alpha^2F(\nu) = {1 \over 2\pi} {\omega_P^2 \over 4\pi}
{d^2 \over d\nu^2} \biggl\{ \nu Re{1 \over \sigma(\nu)} \biggr\}.
\label{explicit}
\ee
\noindent This equation makes it clear that the extraction of $\alpha^2F(\nu)$
by optical measurements is a difficult task. The data must be exceedingly 
accurate as two derivatives are required. Furthermore, the real and imaginary
parts of the optical conductivity must themselves be obtained through
Kramers-Kronig integrations of the reflectance (at least in the case
of reflectometry). This process requires extrapolations at low and
high frequencies. Finally, the plasma frequency must be obtained,
either through a convenient sum rule \cite{kubo}, or from an
independent measurement. In practice we anticipate that accuracy of
reflectance data will limit this inversion technique to qualitative
features of the spectral function. This latter observation further
justifies some of the approximations made above concerning the
neglect of vertex corrections, as they are anticipated to affect the
result only quantitatively.

It is instructive to determine the qualitative accuracy of Eq. (\ref{explicit})
by comparing the result obtained with the true underlying spectral
function. To this end we have carried out the following theoretical
exercise: the conductivity for Pb has been evaluated using Eq. (\ref{sigma}),
starting with the McMillan-Rowell tunneling density of states, and then
Eq. (\ref{explicit}) is utilized to provide $\alpha^2F(\nu)$. We have used
several impurity scattering rates to illustrate that the result is largely
independent of impurity scattering rate, as Eq. (\ref{explicit}) implies.
These results are plotted in Fig. 1.
Also plotted are two curves which are essentially indistinguishable: one is
the $\alpha^2F(\nu)$ used in the 
calculation, the other is obtained through
a numerical inversion of Eqs. (\ref{sigma}) and (\ref{selfenergy}), without
making the weak coupling approximations that led to Eq. (\ref{explicit}).
The details of this more rigorous procedure are described elsewhere
\cite{marsiglio_unpublished}. We include the result here to show that
the procedure works, although the problem is ill-defined (because of the
need to unravel two integrations in Eq. (\ref{sigma}) and (\ref{selfenergy})).
We also remark that the accuracy requirements for the data are quite
stringent, so that in the case of real reflectance data numerical inversion
is problematic. Fig. 1 illustrates that the explicit expression 
(\ref{explicit}) is an excellent approximation to the true electron-phonon
spectral function, particularly since Pb is a fairly
strongly coupled metal ($\lambda = 1.5$). We furthermore learn that using
Eq. (\ref{explicit}) will lead to unphysical negative components of
$\alpha^2F(\nu)$ at high frequency, but that these should essentially
be ignored (the numerically inverted result does not contain these negative
pieces, nor the spurious positive tails). It is also instructive to
examine the effect 
\narrowtext

\begin{figure}
\mbox{
\epsfig{figure=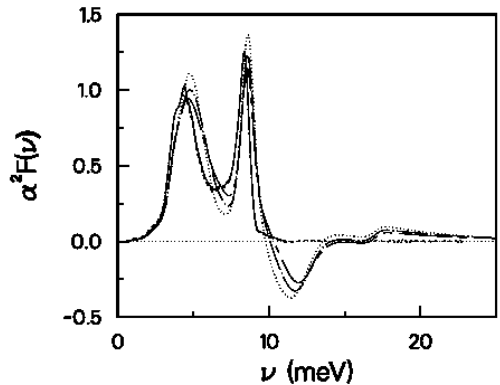,width=0.4\textwidth,clip=}}
\caption{The $\alpha^2F(\nu)$ for Pb (solid curve) plotted vs.
frequency, along with the estimates obtained from Eq. (\ref{explicit})
with an impurity scattering rate, $1/\tau = $ (a) 1 meV (dotted),
(b) 10 meV (dot-dashed), and (c) 20 meV (short-dashed-long-dashed).
Note that all are qualitatively quite accurate, before they
become negative at higher frequencies. Also plotted is the result
(dashed curve, barely discernible from the solid curve)
obtained from a full numerical inversion, as described in the text.}
\label{fig1}
\end{figure}
\narrowtext

\begin{figure}
\mbox{
\epsfig{figure=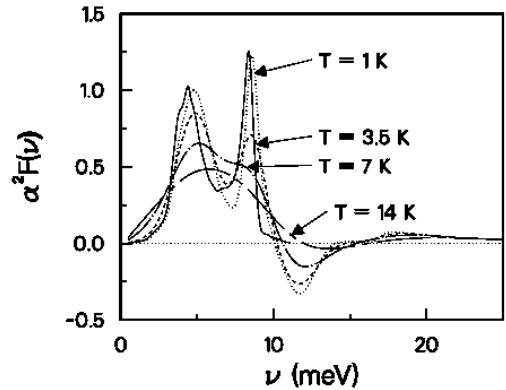,width=0.4\textwidth,clip=}}
\caption{
The $\alpha^2F(\nu)$ for Pb (solid curve) plotted vs.
frequency, along with the estimates obtained from Eq. (\ref{explicit}),
using $1/\tau = 10$ meV, calculated at (a) $T = 1$, (b) $T = 3.5$,
(c) $T = 7$, and (d) $T = 14$ K. As the temperature is raised, the fine
details are lost, but the overall energy scale remains.
}
\label{fig2}
\end{figure}

of finite temperature, especially since we are examining
a normal state property, and the normal state is often inaccessible
at low temperatures (because of superconductivity, for example). In Fig. 2
we show the result of applying Eq. (\ref{explicit}) to the conductivity
for Pb calculated at a series of finite temperatures. The result illustrates
that at sufficiently high temperature the structure is lost, {\it although
the qualitatively correct result remains: that structure in the
conductivity is due to inelastic scattering in the phonon region } (in
the case of Pb). This procedure works fairly well, even though Eq.
(\ref{explicit}) does not even follow in the weak coupling limit at finite
temperature.
With these limitations in mind, we proceed for 
the remainder of this paper
using the explicit equation (\ref{explicit}).
\par

We begin by using the reflectance data at 25 K for K$_3$C$_{60}$ taken by
Degiorgi {\it et al.} \cite{degiorgi}. We have performed our
own Kramers-Kronig analysis on this data, to extract the complex conductivity.
We should caution the reader that this procedure requires low frequency
and high frequency extrapolations. Our results are essentially insensitive to
the form of the high frequency extrapolation.
The standard procedure for the low frequency regime is to use
the Hagen-Rubens form for the reflectance \cite{wooten}; however, we
have used the full Drude form, since second derivatives of the conductivity
are required, and hence the result may be sensitive to these approximations.
We find (see below) that the low frequency result in particular is indeed
sensitive to our choice of low frequency extrapolation.\par

To proceed with a low frequency Drude extrapolation, two parameters are
required, the dc resistivity (as in the Hagen-Rubens form), and an effective
electron scattering rate, which we take to be independent of frequency
for low frequencies. Once these are chosen (by matching the reflectance
with the experimental result at some low frequency), the frequency-dependent
phase can by determined by a Kramers-Kronig integral \cite{wooten}.
From these follow all the optical properties of the solid, in particular
the real and imaginary parts of the conductivity. Equation (\ref{explicit})
then provides us with an estimate for $\alpha^2F(\nu)$. The result is plotted
in Fig. 3, along with the neutron scattering data from \cite{pintschovius}.
We note several features in the figure. First, there are indeed high frequency
negative regions, which our experience with Pb teaches us to ignore. Second,
there is a prominent low frequency negative region (near 20 meV) for which
we have no explanation, except that perhaps the low frequency extrapolation
may be in error. Also note the very low frequency peak at about 5 meV, which
is actually in agreement with the neutron data. On the other hand, an analysis
using the Hagen-Rubens low frequency form (which is {\it less} accurate
than the Drude form) leads to essentially the same result {\it minus} the
low frequency peak. Hence, we cannot say for sure that this peak (attributed
to librons) is coupled to the electrons. The negative region
just above the libron
peak persists regardless of the method of low frequency extrapolation;
aside from the point noted above, it could be due to varying electron
density of states effects \cite{mitrovic}, for example. We proceed, assuming
the problem can be resolved by including such corrections, and discard
the negative pieces in Fig. 3 for the remainder of the paper.
\par

If we ignore the negative pieces of $\alpha^2F(\nu)$ the qualitative
agreement with the neutron scattering data (plotted with a dashed curve
\cite{pintschovius} ) is striking. Not only is the energy scale
correct, but peaks line up fairly well, suggesting that the coupling
is more or less energy independent. 
It is difficult to assess whether contributions are truly coming from
the higher energy region, as the neutron scattering data ends around
200 mev.
\par

In any event, we can think of the $\alpha^2F(\nu)$ in Fig. 3 as an
"educated guess" and now proceed to calculate the expected reflectance.
One might first try to fit the data by selecting a suitable plasma frequency
and frequency-independent (impurity) scattering rate, so that the
conductivity is Drude-like. Such an attempt fails, and it is clear that
inelastic scattering is required. We use Fig. 3 {\it without adjustment}
and vary (by hand) the plasma frequency and impurity scattering rate to
obtain reasonable agreement with the measured reflectance. 
The result is plotted in
Fig. 4. While significant changes in $\alpha^2F(\nu)$
\narrowtext
\begin{figure}
\mbox{
\epsfig{figure=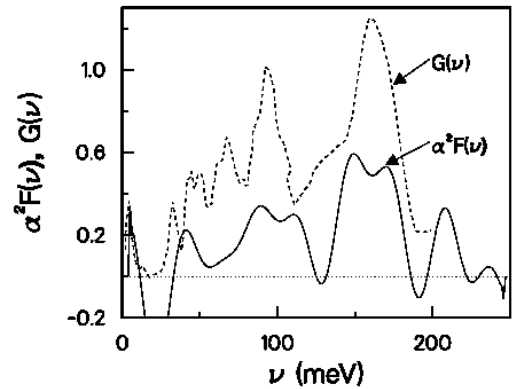,width=0.4\textwidth,clip=}}
\caption{
 The  $\alpha^2F(\nu)$ for K$_3$C$_{60}$ (solid curve)
extracted from the reflectance data of Degiorgi {\it et al.} 
\protect\cite{degiorgi},
using Eq. (\protect\ref{explicit}). For purposes of analysis we have omitted
the negative parts (see text). Also shown are the neutron scattering
results from Ref. \protect\cite{pintschovius} (dashed curve). Clearly the energy
scale in  $\alpha^2F(\nu)$ matches that of the phonons, and some of the
peaks even line up correctly.
}
\label{fig3}
\end{figure}
\narrowtext
\begin{figure}
\mbox{
\epsfig{figure=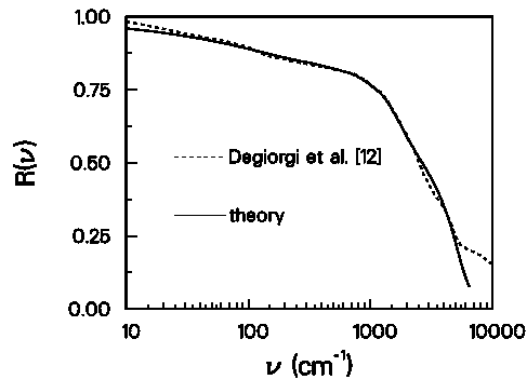,width=0.4\textwidth,clip=}}
\caption{
 The reflectance calculated using the spectral function
from Fig. 3 (solid curve) along with the drude parameters $\omega_P = 6000$
cm$^{-1}$ and $1/\tau = 95$ meV. The reflectance data of Degiorgi
 {\it et al.} \protect\cite{degiorgi} (dashed curve) are also shown.
}
\label{fig4}
\end{figure}

will lead to a
deterioration of the agreement shown in Fig. 4, minor adjustments can
be imposed which will actually improve the agreement. However, this
latter fine-tuning violates the spirit of Eq. (\ref{explicit}), which,
as we learned from the exercise with Pb, provides a qualitatively correct
estimate for the bosonic spectral function.\par

We should note that we obtain as fitted parameters the result $\omega_P
= 6000 $ cm$^{-1}$ and $1/\tau = 95$ meV for the plasma frequency and
impurity scattering rate, respectively. The first result is somewhat less
than the result obtained via the sum rule \cite{kubo}, which was used
in Eq. (\ref{explicit}) ($\omega_P = 7200$ cm$^{-1}$). We view this as
a rather minor inconsistency, compared to, say, that which exists
between the penetration depth-derived plasma frequency, and the one
from the optical sum rule \cite{degiorgi}. The value obtained for the
impurity scattering rate is a little more disconcerting, since it is quite
large. In the approach of Ref. \cite{degiorgi} such a large scattering
rate is avoided by explicitly accounting for a mid-infrared band. 
We could, of course, also include a mid-infrared band with extra parameters
to improve the overall fit and achieve a more realistic scattering
rate, but then the derived $\alpha^2F(\nu)$ would have more model
dependence then is already implied in Eqs. (\ref{sigma}) and
(\ref{selfenergy}).
\par

Given the extracted  $\alpha^2F(\nu)$ in Fig. 3 (without the negative
components) one can ask whether or not such a spectral function can account
for the superconducting properties in K$_3$C$_{60}$. To this end we have
computed the Coulomb repulsion pseudopotential, $\mu^\ast$, given that $T_c
=19$ K. The result is  $\mu^\ast = 0.4$ (using a cut-off, $\omega_c = 1$ eV). 
The spectrum can be characterized by $\lambda = 1.2$ and $\omega_{ln} =
40$ meV. This gives us $T_c/\omega_{ln} = 0.04$, which, in spite of the
rather large value of $\lambda$, represents a fairly weak coupling strength.
Nonetheless, the gap ratio, $2\Delta_0/k_BT_c$ comes out fairly close
to 4, significantly different from the BCS weak coupling result, a result
which is partially due to the large value of $\mu^\ast$ \cite{sammer}.
If the low frequency peak in $\alpha^2F(\nu)$ at 5 meV is excluded, the
results are then {\it very} BCS-like: $\lambda = 0.8$, $\mu^\ast = 0.34$,
$T_c/\omega_{ln} \approx 0.016$, and the gap ratio is
$2\Delta_0/k_BT_c \approx 3.6$. Either of these possibilities is
strongly refuted by a gap measurement of order 5 \cite{zhang}, though
it is difficult to discriminate between the two lower values
based on far-infrared
measurements. Microwave measurements may provide a discriminating probe,
as discussed in \cite{marsiglio97}. 
\par

In conclusion, we have described an inversion scheme to extract the
source of inelastic scattering in a metal. Such a scheme has many
advantages over that already utilized in tunneling, not the least of which
is Eq. (\ref{explicit}), which allows a direct estimate from optical
measurements with essentially
no effort. Fairly accurate experiments are required however. We applied
this technique to K$_3$C$_{60}$, and have found the {\it qualitatively correct}
$\alpha^2F(\nu)$ for this material. The coupling is sufficiently strong
to explain the superconductivity in this material; in fact the $\mu^\ast$
required is conspicuously large, so that, as suspected by many, 
electron-electron correlations are strong as well. Nonetheless, the weak
coupling approach presented here appears to be able to account for
the main features observed in the far-infrared. K$_3$C$_{60}$ is a
conventional weak coupling electron-phonon superconductor with an
anomalously large Coulomb pseudopotential.\par 

We acknowledge helpful discussions with Tom Timusk. This research was supported
by the Natural Sciences and Engineering Research Council (NSERC) of Canada and
by the Canadian Institute for Advanced Research (CIAR).

\end{multicols}

\begin{thebibliography} {999}
\bibitem{remark_allen}
We perform the weak coupling expansion somewhat differently from
Allen \cite{allen71}. The denominator $\nu + {i \over \tau}$ is kept
as the unperturbed part; once the integral in the expanded portion
 is performed, we de-expand (as Allen does---see his Eq. (61))
but we do not attempt to extract a renormalization component which
leads to his undetermined function $\Phi(\omega)$. Our result has
no unknown components.
\end{thebibliography}
\end{document}